\documentclass[11pt,onecolumn]{article}

\usepackage{graphicx}
\usepackage{dcolumn}
\usepackage{bm}
\usepackage{amsmath}
\usepackage{amssymb}
\usepackage{amsfonts}
\usepackage{siunitx}
\usepackage{xcolor}
\usepackage{physics} 
\usepackage[normalem]{ulem}
\usepackage[bottom]{footmisc}
\usepackage{soul}
\usepackage{float}
\usepackage{cancel}
\usepackage{url}
\usepackage[colorlinks=true,citecolor=blue,urlcolor=red]{hyperref}
\usepackage{pdfpages}

\begin{document}
	\begin{center}	
		{\bf{ {Experimental realization of
					an analog of entanglement between two Brownian particles}}}
		
		%
		{Lakshmanan Theerthagiri  $^{1, 2}$, Sergio Ciliberto$^{3}$}

		{\small $^1$ Physics Division,  University of Camerino, I-62032 Camerino (MC), Italy}
		
		{\small $^2$Department of Physics, University of Naples ``Federico II", I-80126 Napoli, Italy}
		
		{\small $^3$ Univ Lyon, ENS de Lyon,  CNRS, Laboratoire de Physique, F-69342 Lyon, France}
	\end{center}

{\bf Abstract:}  
We experimentally investigate the statistical properties of  {a classical analog of quantum entanglement} considering  two Brownian particles connected by an elastic force and maintained at different temperatures through separate heat reservoirs. Uncertainty relations between coordinates and coarse-grained velocity can produce a phenomenon similar to quantum entanglement, where temperature plays the role of Planck's constant.  The theoretical analysis matches the experimental results, confirming that the interconnected particles exhibit Brownian quantum-inspired classical correlation entanglement. This effect arises from a coarse-grained description of Brownian motion and vanishes at a finer resolution. {The coarsening scales range is measured too.}		
\vspace{1cm}

	The search of classical analogies of quantum mechanical phenomena is an old and widely studied  problem going back to the origin of quantum mechanics. Indeed quantum mechanics and classical statistical physics share probabilistic foundations, leading to inherent similarities. 
	
	Analogies can be drawn between quantum complementarity and aspects of statistical thermodynamics, such as the energy-temperature uncertainty relation and links between  Fokker-Planck and Schrödinger equations illustrate the interplay between stochastic and quantum descriptions. \cite{nelson,furth}. Attempts have been made in classical kinetic models to attain the Schrödinger equations and stochastic electrodynamics \cite{Risken,millard,KANIA}.
	{This is   relevant to other fundamental physical problems, such as a quantum particle coupled to a quantum
	heat bath—a key concept in statistical mechanics, condensed matter, and quantum optics. In the classical limit
	($\hbar \rightarrow 0$), using the correct formula \cite{ford} one recovers the familiar
	white-noise result  and  physically consistent Langevin and Fokker-Planck equations \cite{Ford_PRA,Lutz2025} }

	 
	{ When examining the link between quantum mechanics and classical statistical physics, one may ask whether quantum phenomena like entanglement—central to quantum optics, information tasks \cite{warit}, and quantum computing—have classical counterparts. In this context, coarse-grained measurements, decoherence \cite{Preskill}, and collapse models have been widely studied \cite{brukner,chaudhary,mal} to explain the quantum-to-classical transition.}
	
	{To give more insight into these problems we describe in this letter the results of an experiment on two interacting Brownian subsystems (particles) in contact with two thermal baths. We performed this experiment to study the possibility of entanglement between  coarse grained observables as a function of the coarsening  time scale $\Delta$.{ We define the coarse grained velocity $V$ as the particle velocity on the scale  $\Delta$.  We show that uncertainty relations exist between positions and coarse grained velocities in a wide range of $\Delta$. }From these uncertainty relations we define and  measure quantum inspired inequalities  that are sufficient conditions for the entanglement between two interacting Brownian particles \cite{furth,all} showing the analogy to the continuous variables quantum entanglement. The thermal bath temperature acts like  Planck's constant in the uncertainty relations between coordinate and coarse grained velocity fluctuations of the classical system, leading to a phenomenon akin to quantum entanglement.}
	 {Throughout the letter either "Brownian" or "classical" entanglement has to be understood as a classical analogy of its quantum counterpart}. 
	
	In the past other studies  on ''classical entanglement'' and coarse graining have been developed. Classical analogies to quantum entanglement are  found  
	 in quantum optics, where the non-quantum entanglement resolves basic issues in polarization optics \cite{Rsimon} and in wave physics where the  analogy is used to get coherent imaging from two beams \cite{Lugiato_2004}.
	 { Experiments on macroscopic quantum coherence have been proposed \cite{leggett,hardy}
	and in the context of measurements, both theoretical \cite{Roh, mukherjee} and experimental \cite{AJ,Friedman} frameworks  have been designed for coarse-grained measurement scenarios \cite{kim,prasanta}.} {None of these previous studies deal with the influence of coarse graining scale on the classical entanglement that we discuss in this letter} 
	
	\begin{figure}
		\centering
		\includegraphics[width=0.35\textwidth]{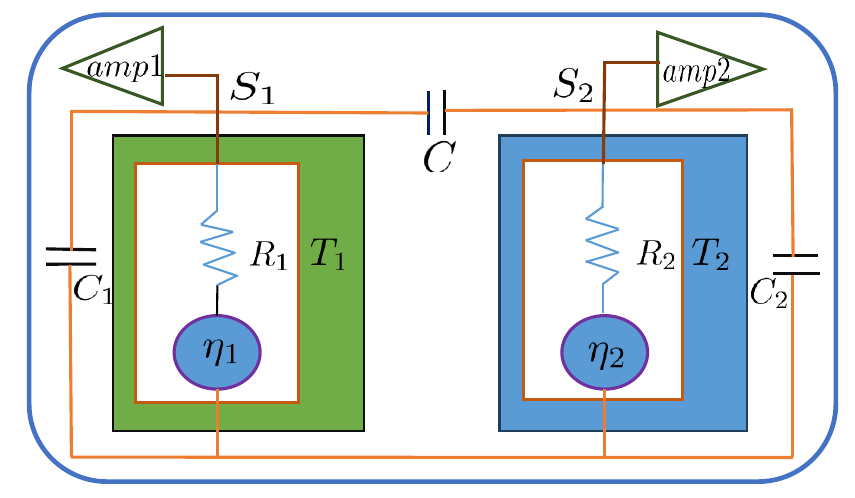} \\ \includegraphics[width=0.35\textwidth]{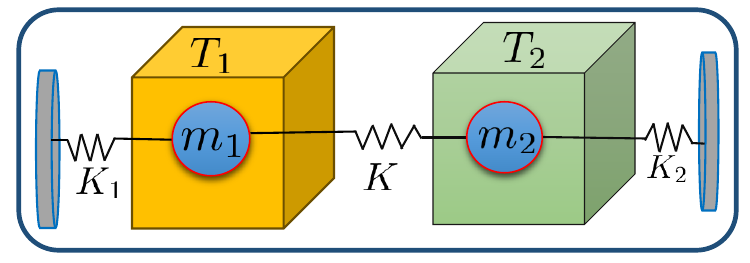} 
		\caption{(a) Diagram of the circuit. The resistances $R_1$ and $R_2$ are kept at temperature {$T_1=$ [$88$K-$300$K] and $T_2=296$K},
			respectively. They are coupled via the capacitance $C$. The
			capacitances $C_1$ and $C_2$ schematize the capacitance of the cables
			and of the amplifier inputs. The voltages $S_1$ and $S_2$ are amplified
			by the two low noise amplifiers $amp_1$ and $amp_2$. (b) The circuit in
			(a) is equivalent to two Brownian particles ($m_1$ and $m_2$) moving
			inside two different heat baths at $T_1$ and $T_2$. The two particles are
			trapped by two elastic potentials of stiffness $K_1$ and $K_2$ and
			coupled by a spring of stiffness $K$. 
			The corresponding mathematical model is defined in Eq.\ref{lan}.
			}
		\label{circuit}
	\end{figure}

	Our experimental setup, shown in Fig.~\ref{circuit}, comprises two resistors, labeled as $R_1$ and $R_2$. These resistors are kept at different temperatures, $T_1$ and $T_2$ respectively. While $T_2$ remains constant at $296K$, we have the flexibility to adjust the temperature of $T_1$ between $296K$ and $88$ $K$ by using a layered vapor setup over a liquid nitrogen bath. The Fig.~\ref{circuit} illustrates the resistors along with their associated thermal noise generators, denoted as  $\eta_1$ and $\eta_2$, whose power spectral densities follow the Nyquist formula: $\vert \Tilde{\eta}_j\vert^2 = 4k_BR_jT_j$, where $j=1,2$.

	The energy exchanged between the two thermal baths is governed by the coupling capacitance $C$. $C_1$ and $C_2$ indicates the circuits and cables capacitance, {kept at $T_2$}.  All the relevant quantities considered in this study can be derived from the measurements of the voltages $S_n$ ($n=1,2$) across the resistors $R_n$ (see Ref.\cite{Ciliberto_2013,supmat} for details).
	

	 Circuit analysis reveals the following equations describing the charges dynamics:
	\begin{equation}\label{lan}
		\begin{split}
			R \dot{q}_1 & =-A_1 q_1+ K \ q_2+\eta_1 \\
			R \dot{q}_2 & =-A_2 q_2+K \ q_1 +\eta_2
		\end{split}
	\end{equation}
	where $R=R_1=R_2$,
	$ A_1=K_1+K,A_2=K_2+K$ with $ K_1=\frac{C_2}{D}, K_2=\frac{C_1}{D},  K=\frac{C}{D}$ and $D=C_1C_2+C(C_1+C_2)$
	Here,  $\eta_j$ represents a delta correlated  noise with properties $\langle \eta_i(t)\eta_j(t')\rangle=2\delta_{ij}k_BT_iR_j\delta(t-t')$ \cite{ford},where $k_B$ is the Boltzmann constant. 
	
This system is governed by the same equations as two Brownian particles maintained at different temperatures and connected by an elastic force as shown in Fig.\ref{circuit}b). In such a case 
	$q_j$ signifies the displacement of particle $j$, $\dot q_j$ represents its velocity, $K_j$ denotes the stiffness of spring $j$, $K$ that of   the coupling spring, and $R$ the viscosity. Furthermore Eqs.\ref{lan} correspond to those of the Brownian gyrator 
	\cite{Gyrator_1,Gyrator_2,Gyrator_3}
	which is one of the simplest out of equilibrium models of heat transfer.

	We expand the use of this experimental setup to include theoretical studies exploring Brownian entanglement in coupled harmonic oscillators in the over-damped regime \cite{all,pooja}. This occurs when the relaxation time of the coordinate
	$\tau_x$ significantly exceeds that of the momentum 
	$\tau_p$. When considering timescales $\tau_x\gg \tau_p$, the focus is on the changes in the coordinate, leading to the definition of the coarse-grained velocity
	$v=\Delta x/\Delta t$ for $\tau_x\gg \Delta t\gg \tau_p$.

	In order to properly define the  uncertainty relations that we want to study in the experiment let us first recall several useful theoretical results. In the quantum case following Ref.\cite{all} one can propose  a simple sufficient condition for entanglement. Let us consider two harmonic oscillators of mass $m$ and  frequency $\omega$ with coordinates and momenta operators $\hat x_i$ and $\hat p_i$ with $i=1,2$
	We first note, that for 
	$$ \Delta\hat{x}_i=\hat x_i-<\hat x_i> \ \ \ \rm{and} \ \ \  \Delta\hat{p}_i=\hat p_i-<\hat p_i>,$$
	the standard uncertainty relations  \cite{furth}
	\begin{eqnarray}\label{uncer}
		\langle \Delta\hat{x_i}^2\rangle  \langle \Delta\hat{p_i}^2\rangle\geq\frac{\hbar^2}{4} 
	\end{eqnarray}
	imply \cite{simon,zoller},
	\begin{equation}
		m_i\omega^2 \langle \Delta\hat{x_i}^2\rangle + {1 \over m_i} \langle \Delta\hat{p_i}^2\rangle\geq  \langle m_i\omega^2  \Delta\hat{x}^2\rangle + {1 \over m_i} \frac{\hbar^2}{4 \langle \Delta\hat{x}^2\rangle}.
	\end{equation}
	
	The right hand  side has a minimum at $ \langle \Delta\hat{x}^2\rangle =\hbar/(2 m_i\omega)$, thus one gets:
	\begin{equation}\label{varia}
		m_i\omega\langle \Delta\hat{x_i}^2\rangle + {1 \over m_i\omega}\langle \Delta\hat{p_i}^2\rangle\geq  \hbar.
	\end{equation}
{  If the two particles are in a separable state—meaning that their joint probability distribution for coordinates and momenta can be expressed as a statistical mixture of product states—then the corresponding variances satisfy the following additive relation \cite{holger,all}} :
	\begin{eqnarray} \label{eq:notentangled}
		& & m_i\omega \langle (\Delta x_1-\Delta x_2\rangle)^2\rangle + {
		\langle (\Delta p_1+\Delta p_2\rangle)^2\rangle \over m_i\omega}= \ \ \ \ \notag\\
		&=& m_i\omega  \left(\langle\Delta x_{1}^2\rangle+\langle\Delta x_{2}^2\rangle \ \right)
		+{  \langle\Delta p_{1}^2\rangle+\langle\Delta p_{2}^2\rangle  \over m_i\omega}\geq 2\hbar.
	\end{eqnarray}
	
{	The  violation of this uncertainty limit  
	 proves that the quantum state cannot be separated into a mixture of product states and that 
	\begin{eqnarray}\label{condition}
		m_i\omega\langle (\Delta x_1+\epsilon\Delta x_2\rangle)^2\rangle+
		{ \langle (\Delta p_1+\zeta\Delta p_2\rangle)^2\rangle \over m_i\omega} \le 2\hbar,
	\end{eqnarray}
for at least  one of four independent choices $\epsilon=\pm 1$ and $\zeta=\pm 1$, is a sufficient condition for entanglement \cite{holger}.}

	In order to find similar uncertainty relations for the observables of our experimental system we see that Eq.~\eqref{uncer} necessitates incorporating momenta. However the velocity in an underdamped system is unbounded and we have to properly define a coarse-grained velocity $V$, using experimental coordinates $X_i=q_i$ ($i=1,2$). This results in two separate velocities. For the  ensemble of all trajectories $\bold{X}=(q_1,q_2)=(x_1,x_2)$  \cite{all}, the average coarse-grained velocity $V_n$ of the Brownian particle $n=1,2$  can be  defined as
	\begin{equation}  \label{eq:coarse_vel_a}
		V_{\pm,n}(\bold{X,t})=\lim_{\Delta\rightarrow 0 } V_{\pm,n}(\bold{X},t,\Delta),
    \end{equation}
	where 	
		\begin{equation}\label{eq:coarse_vel_b}
		V_{\pm,n}(\bold{X},t,\Delta)=
		\int d\bold{y}\ \frac{(\pm y_n\mp x_n)}{\Delta}\ P(\bold{y},t\pm \Delta\vert \bold{X},t),
	\end{equation}
and  $P(\bold{y},t\pm \Delta\vert \bold{X},t)$ is the conditional probability of being in $y$ at time $t\pm\Delta$ if the system is in $X$ at time $t$. We also assume that $\Delta$ is much larger than the (real) momentum's characteristic relaxation time $\tau_p$  which is small in the overdamped limit.\\

	Since regular trajectories are absent, it is necessary to define different velocities for different time directions, getting the following theoretical predictions ( see ref.\cite{all} and SM\cite{supmat})
	\begin{eqnarray}
			V_{+,n}(\bold{X},t)&=& {f_n(\bold{X}) \over R} \\
		V_{-,n}(\bold{X},t)	&=&{f_n(\bold{X}) \over R}-{2k_BT_n  \over R} \partial_{x_n}\ln{P(\bold{X},t)}
		\label{eq:V+-},
	\end{eqnarray} 	
where $f_n=-\partial_{x_n} U(\bold{X})$ and $U(\bold{X})$ the potential, which for Eqs.\ref{lan} is $U(q_1,q_2)=A_1 q_1^2/2+ A_2 q_2^2/2-K q_1q_2$.
	These velocities have specific physical interpretations: $V_{+,n}(\bold{X},t)$ denotes the average velocity needed to move anywhere  from the point $\bold{X}$, whereas $ V_{-,n}(\bold{X},t)$ indicates the average velocity from any point to reach $\bold{X}$ at a specific time $t$. 
	 Furthermore it  can be shown that the mean $\bar V_n=(V_{-,n}+V_{+,n})/2$ is  related to the probability current $J_n(X,t)$ such that $\bar V_n P(X,t)=J_n(X,t)$ and to the entropy production rate.  \cite{therm_efficiency_2009,supmat}. 
	 
	 Instead the Coarse-Grained-Velocities Difference (CGVD) 	
	\begin{eqnarray}\label{eq:uj}
		u_n(\bold{X},t)&=&\frac{ V_{-,n}(\bold{X},t)- V_{+,n}(\bold{X},t)}{2}= \label{eq:def_u} \notag  \\&=&-{k_B T_n  \over R} \partial_{x_n}\ln{P(\bold{X},t)}.\label{diffcoarse}
	\end{eqnarray}
measures  the time asymmetry $V_{-,n}\ne V_{+,n}$ produced by the action of thermal baths on the dynamics (factor $T_n$  in Eq.\ref{diffcoarse}). The CGVD $u_n$, the fundamental quantity of our study \cite{Comment_1}, has several useful properties. 
	Indeed from Eq.\ref{eq:uj} it can be  shown \cite{all,supmat} that $\langle q_n \rangle=\langle u_n \rangle=0$ and most importantly that expectation value of the coordinates and of the  CGVD (Eq.\ref{diffcoarse}) are related by \cite{furth,all} :  
	\begin{equation} \label{eq:qnvn}
		\langle q_n u_m \rangle={k_B T_n \over R} \delta_{nm}.
	\end{equation}	
	This equation  shows that the velocity  $u_n$ correlates only with its own coordinate $q_n$. This is due to the fact that the thermal baths acting on different Brownian particles are independent of each other. By applying the standard Cauchy-Schwartz inequality to Eq.~\eqref{eq:qnvn} \cite{all}, {we can deduce  an  uncertainty relation:}
	
	\begin{equation}\label{eq:uncert}
		\begin{split}
			\langle q_m^2\rangle\langle u_n^2\rangle \geq  \vert \langle q_m u_n \rangle \vert^2
			={\left( k_BT_n \over R \right)^2 } \delta_{mn},
		\end{split}
	\end{equation}
	which is formally similar to Eq.\ref{uncer}, with $T_n$ playing the role of $\hbar$. 
		Proceeding as for the quantum case we can write : 
	\begin{equation}
		\langle q^{2}_{j}\rangle+\frac{R^2}{A^{2}_{j}}\langle u^{2}_{j}\rangle\ > \langle q^{2}_{j}\rangle+\frac{R^2}{A^{2}_{j}}  {\left( k_BT_j \over R \right)^2 {1\over \langle q^{2}_{j}\rangle }}.
	\end{equation}
	The right hand side has a minimum at $\langle q^{2}_{j}\rangle= {k_BT_j \over A_j}$ thus 
	\begin{equation} \label{eq:ineq}
		{A_{j}}  \langle  q^{2}_{j}\rangle+\frac{R^2}{A_j}\langle u^{2}_{j}\rangle\ >\ {2 \ k_BT_j}.
	\end{equation}
	We can use the coupled system to obtain the Brownian entanglement relation, as  expressed in Eq.\eqref{condition} for the quantum case : 
	\begin{equation}\label{explong}
		\begin{split}
			&\langle \left(q_1\sqrt{A_1}+\epsilon q_2  \sqrt{A_2}\right)^2\rangle+ \\ & + R^{2}\langle\left(\frac{u_1}{\sqrt{A_1}}+\zeta \frac{u_2}{\sqrt{A_2}}\right)^2\rangle\ < \ 2k_B(T_1+T_2).
		\end{split}
	\end{equation} 
	This confirms that for values of $\epsilon= \pm 1$ and $\zeta=\pm 1$, in at least one of the four independent choices that meet both the entanglement and uncertainty conditions, it becomes possible to create entanglement between two Brownian particles.{ Uncertainty relations Eqs.\ref{uncer}, \ref{eq:uncert} with  Eqs.\ref{condition},\ref{explong}, establish a sufficient condition for entanglement. Key quantities involved are momentum and coordinates in the quantum case, whereas in the Brownian context, they are coarse-grained velocity $u_i$ and coordinates $q_i$.}

\begin{figure}
		\centering
			\hspace{-2cm}\textbf{a)} \hspace{4cm}  \textbf{b)} \\
		\includegraphics[width=0.23\textwidth]{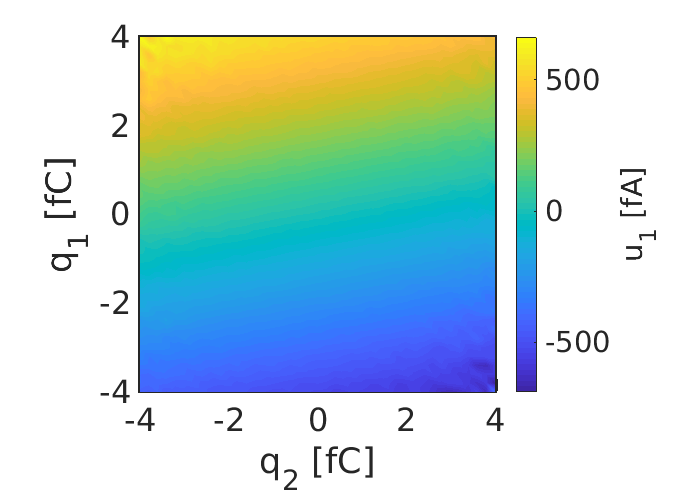}
		\includegraphics[width=0.23\textwidth]{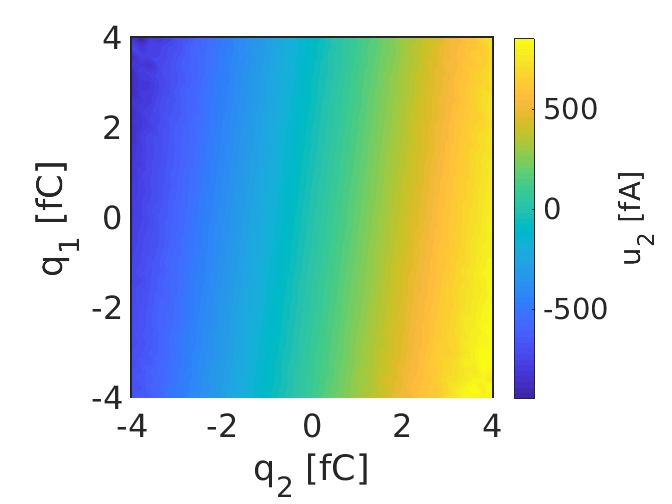} \\
		\includegraphics[width=0.23\textwidth]{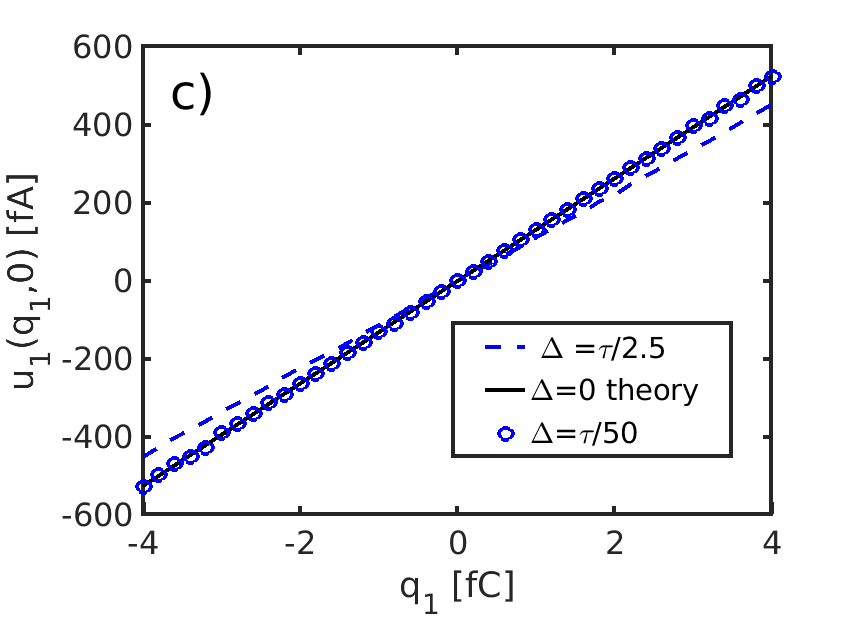}
		\includegraphics[width=0.23\textwidth]{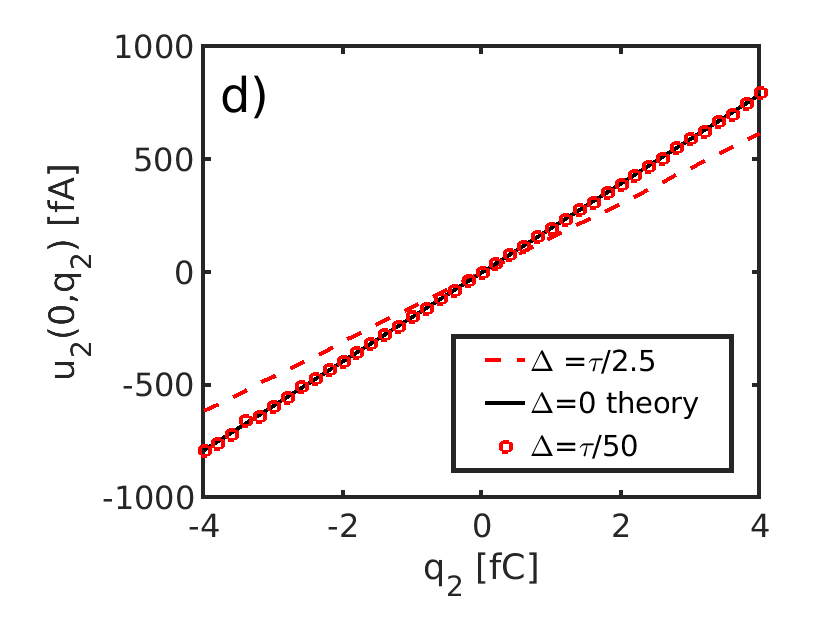}\\
		\caption{a), b) Coarse grained velocities differences as a function of $q_1$ and $q_2$ at $\Delta=\tau/50$. c)  cross section of $u_1$ versus $q_1$ at $q_2=0$ at $\Delta=\tau/50$ {\color{blue}$\circ$} and $\Delta=\tau/2.5$. d)  cross section of $u_2$ versus $q_2$ at $q_1=0$ at $\Delta=\tau/50$ {\color{blue}$\circ$} and $\Delta=\tau/2.5$. The black lines are the theoretical predictions from Eq.\ref{diffcoarse}}
		\label{fig:coarse_grained_vel}
	\end{figure}

	{We start the experimental investigation of the Brownian entanglement(Eq.\ref{explong}) in the isothermal case, $T=T_1=T_2$, by studying how  a finite sampling time $t_s$  affects the results.  Indeed the evaluation of $u_n$ using Eq.\ref{eq:def_u} implies $\Delta \rightarrow 0$  in Eq.\ref{eq:coarse_vel_a}, but   in experiments the smallest time resolution is $2 t_s$ (Nyquist–Shannon theorem), as a consequence   $\Delta\ge 2t_s$. In our  experiment $t_s=122.07\mu$s, which is about $\tau/50$ where  $\tau=2\ R/(A_1+A_2)\simeq 6ms$ is the main relaxation time of the system.	
	Thus we have to check whether this $t_s$ is sufficiently short for a correct estimation of $u_n$.} 
		 In order to do this check  we measure   
		$u_{n,\Delta}= ( V_{-,n}(\bold{X},t,\Delta)- V_{+,n}(\bold{X},t,\Delta))/{2}$ using in Eqs.\ref{eq:coarse_vel_b} the conditional probabilities computed from the experimental time series of $q_1$ and $q_2$ (see SM for details \cite{supmat} ). 
		Fig.\ref{fig:coarse_grained_vel} depicts the  $u_{n,\Delta}$ calculated at $\Delta\simeq \tau/50$.  The functions $u_{n,\Delta}(q_1,q_2)$ exhibit a linear dependence on $q_1$ and $q_2$, although the slopes vary with $\Delta$. This variability is evident in the lower panels of Fig. \ref{fig:coarse_grained_vel}, where we plot cross-sections of $u_{1,\Delta}$ and $u_{2,\Delta}$ at $q_2=0$ and $q_1=0$ respectively, for two different values of $\Delta$. The theoretical estimations for the slopes are recovered only when $\Delta \ll  \tau$.
		
\begin{figure}
		\centering
		\includegraphics[width=0.5\textwidth]{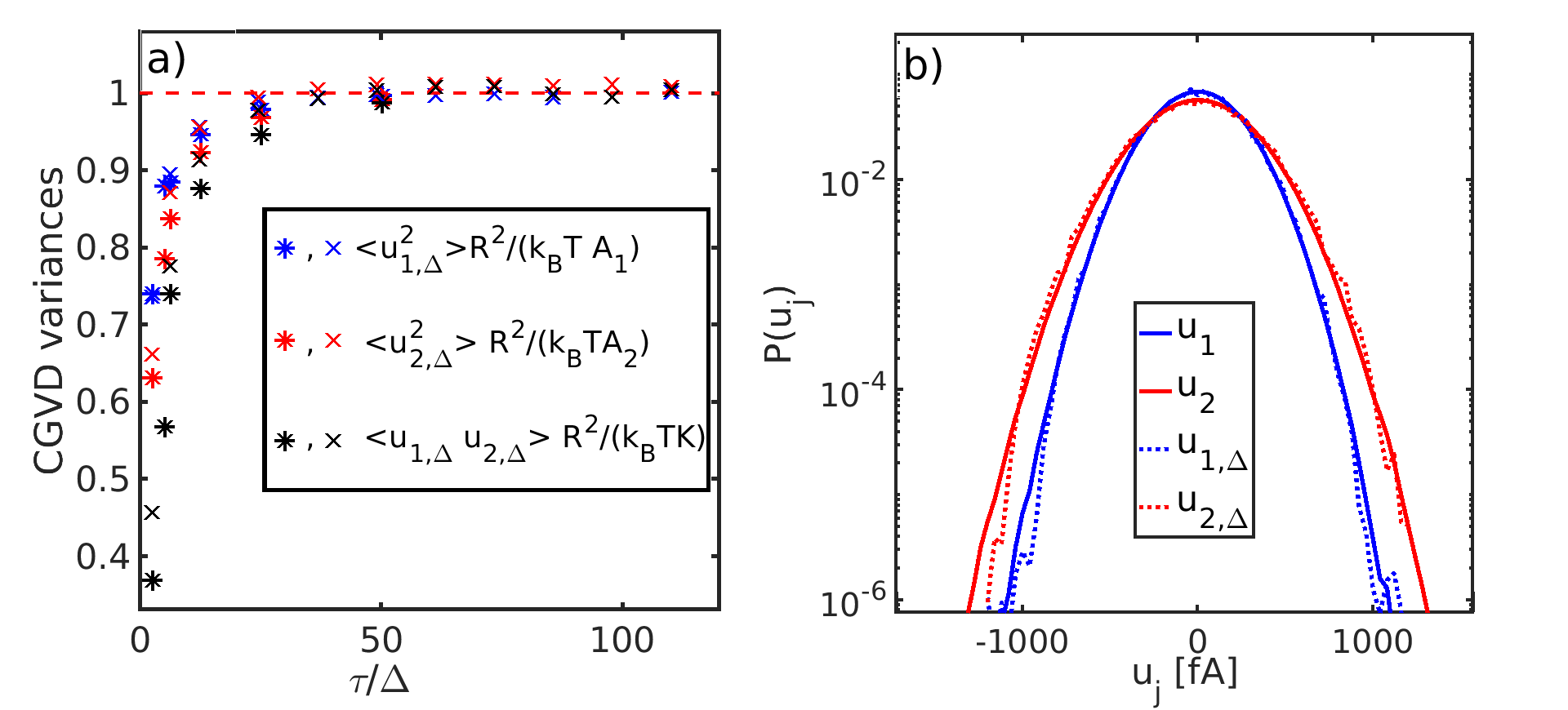}
		\caption{a)  The numerical ($\cross$) and experimental ($\ast$) values of  the measured variances of  $u_{j,\Delta}$   are plotted as a function of $\tau/\Delta$. Blue and red symbols correspond to  j=1 and  j=2 respectively. The black symbols correspond to the cross-variance.  As indicated in the figure the values of the variances and cross variance are normalized  to their respective theoretical predictions at $\Delta=0$ obtained  from Eq.\ref{diffcoarse} (see \cite{supmat}).   
		b) 	Pdf of $u_j$, with j=1 (blue) and j=2 (red). The continuous lines correspond to $u_j$ estimated from Eq.\ref{diffcoarse} whereas the dotted lines correspond to the pdf of  $u_{j,\Delta}$. These ones have larger statistical fluctuations because in order to measure them one needs to estimate the conditional probabilities in Eq.\ref{eq:coarse_vel_b}.  }
		\label{fig:pdf_uj}
	\end{figure}

	{
	This dependence on $\Delta$ of $u_{n,\Delta}$  can be measured by computing their variances  and comparing them to the values directly estimated using Eq.\ref{diffcoarse} under isothermal conditions $T_1=T_2=T$. In such a case the theoretical estimations are: $\langle  u_n^2 \rangle = (k_B T) A_n /R^2$ and $\langle u_n u_{n'}\rangle= (k_B T) K /R^2$ with $n\ne n'$ (see ref.\cite{supmat}). The ratio between  the  variances of $u_{n,\Delta}$ and those of $u_n$ (computed from Eq.\ref{diffcoarse})  are plotted as a function of $\tau/\Delta$ in Fig.\ref{fig:pdf_uj}a). This ratio  is close to 1 within error bars for $\Delta<\tau/40$, confirming the findings from Fig.\ref{fig:coarse_grained_vel}. The results at $\Delta<ts/2$ have  been obtained from a numerical simulation of Eqs.\ref{lan} which uses the  experimental values of the parameters. 
	To strength this results we compare in Fig.\ref{fig:pdf_uj}b) the probability density functions (pdf)   of the fluctuations of  $u_n$ and of $u_{n,\Delta}$  at $\Delta=\tau/50$.  
	The pdf of $u_{n,\Delta}$ and $u_n$ almost coincide except for the large statistical errors on  the pdf of $u_{n,\Delta}$. These errors  are  induced by the estimation of $P(\bold{y},t\pm \Delta\vert \bold{X},t)$ in Eq.\ref{eq:coarse_vel_b} which requires a very large number of data points  to be correctly evaluated. 
	For this plot we used $5 \times 10^7 $ data points and we checked that indeed the fluctuations on the pdf tails increase by reducing the number of data.
	Based on the results on the variances  and on the pdf (Figs.\ref{fig:pdf_uj}a) and b), we conclude that to reliably measure the coarse grained velocities, $t_s$ and 
	$\Delta$ must be smaller than $\Delta<\tau/40$, which holds true in our experiment.}
%
%

	\begin{figure}
		\centering
		\includegraphics[width=0.5\textwidth]{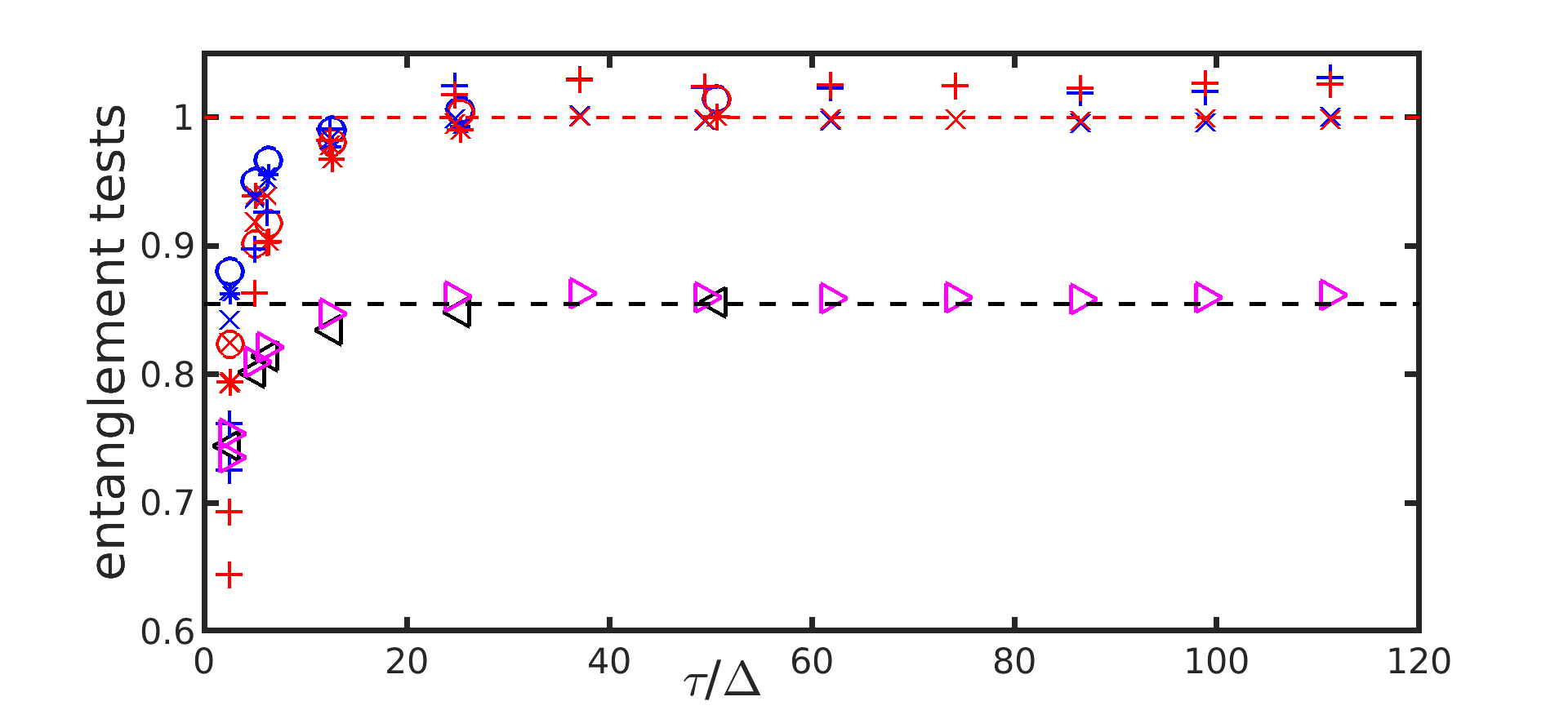}
		\caption{
			Entanglement test versus $\tau/\Delta$.  The `{\large$\times$}' correspond to numerical values and `{\large$\ast$}'  to experimental values of $CR_{n,\Delta}=\langle q_n u_n \rangle/(k_BT/R)$. The `{\large$\circ$}' and the  `{\large$+$}' represent respectively  the experimental and numerical values of the quantity $UI_{n,\Delta}=\langle q_n^2\rangle \langle u_n^2 \rangle /(k_BT/R)$.Blue and red  stand for $n=1$ and $n=2$ respectively.   
			At small $\Delta$, $CR_{n,\Delta}\simeq 1$ and $UI_{n,\Delta} >1 $, i.e. Eq.\ref{eq:qnvn} and Eq.\ref{eq:uncert} are both satisfied. 
			  The experimental ($\triangleleft$) and numerical  values ($\triangleright$) of the entanglement coefficient $EC_\Delta$ saturate showing  that the theoretical prediction holds true within the margin of error for all $\Delta$ values less than $\tau/40$. This confirms the classical entanglement criteria of Eqs.\ref{eq:uncert} and \ref{explong}
		}
		\label{fig:entanglement}
	\end{figure}

{We can now test  the correlation relation (CR) and the  uncertainty inequalities (UI)  as a function of $\tau/\Delta$. As for large $\Delta$ the coarse grained  velocity are  not well defined, we first check whether Eq.\ref{eq:qnvn} and Eq.\ref{eq:uncert} are  satisfied. This can be seen in  Fig.\ref{fig:entanglement} where we plot $CR_{n,\Delta}=\langle q_n u_{n,\Delta}\rangle  R/k_B/T_n$ (from Eq.\ref{eq:qnvn}) and $UI_{n,\Delta}=\langle q_n^2\rangle \langle u_{n,\Delta}^2 \rangle (R/ k_B/T_n)^2$ (from Eq.\ref{eq:uncert}) versus $\tau/\Delta$.
We clearly see that, for $\Delta < \tau/40$, $CR_{n,\Delta}$ and  $UI_{n,\Delta}$ are constant within error bars and  most importantly they reach the values  $CR_{n,\Delta}\simeq 1$ and $Ur_{n,\Delta}\simeq(1+C^2/D)>1$ which are those theoretically predicted. This means that Eq.\ref{eq:qnvn} and  Eq.\ref{eq:uncert} are satisfied  for all  $\Delta < \tau/40$.  
	
	Finally to estimate entanglement, we define an   entanglement coefficient  $E_c({\Delta}) $ as the ratio between the left and right sides of Eq.\ref{explong}, where $u_{n,\Delta}$ has been used to evaluate the left hand side. The experimental ($\triangleleft$) and numerical ($\triangleright$)  values of the ratio $E_c({\Delta}) $ demonstrate that the theoretical prediction(black horizontal dashed line \cite{supmat}) is achieved within error bars for all $\Delta < \tau/40$, thereby confirming the criteria for classical entanglement, i.e. the inequality Eq.\ref{explong} is satisfied for $\Delta \rightarrow 0$. The error bars, which are of the order of the symbol size, primarily stem from statistical errors in both numerical simulations and experiments, as well as residual calibration inaccuracies in the experiment.
	\begin{figure}
		\centering
		\includegraphics[width=0.23\textwidth]{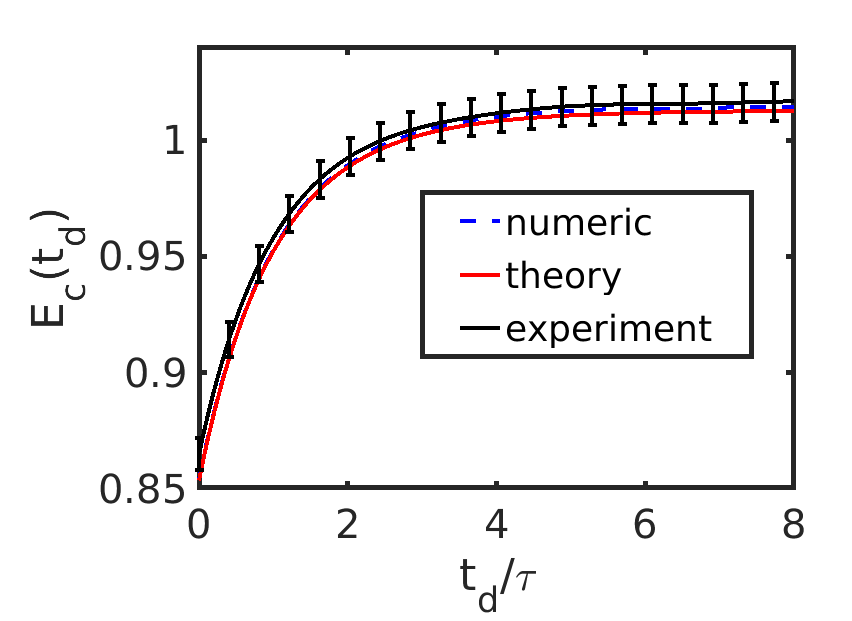}
		\includegraphics[width=0.23\textwidth]{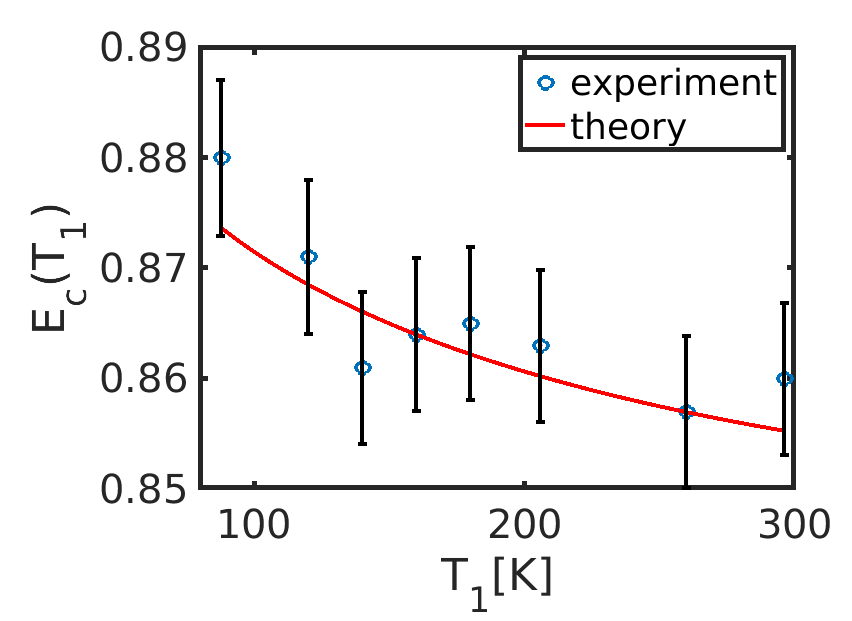}
		\caption{a) $E_c$ versus the delay time  $t_{d}$ between the measures of $(q_1,u_1)$ and $(q_2,u_2)$.  The entanglement coefficient is 0.85  at $t_{d}=0$ (as in Fig.\ref{fig:entanglement}), but it rises with $t_{d}$. It surpasses $1$ at about $4t_{r}$, meaning that entanglement is lost as Eq.\ref{explong} is no longer satisfied. The theoretical prediction, represented by the red line, is aligned with the experimental black line and numerical values blue line. b) $Ec$ and the theoretical prediction as a function of $T_1$. The maximum entanglement, or the lowest entanglement coefficient, occurs under isothermal conditions, and the entanglement coefficient is barely affected by changes in $T_1$.	}
		\label{entanglement_versus_delay}
	\end{figure}

	Of course the measured entanglement must disappear after a certain time if we instantaneously set $K=0$, i.e. the  particles interaction is switched off. However this test cannot  be realized experimentally because of the  unavoidable noise introduced by the commutation. Thus we use the alternative approach of the delayed measurements where the variables are measured with a delay time $t_{d}$.} As we are in a stationary state this delay will not affect the variances $u_j$ and $q_j$, but it will affect the cross variances $\langle u_1(t_{d}) u_2(0)\rangle $ and  $\langle q_1(t_{d}) q_2(0)\rangle $, which of course decrease for increasing $t_{d}$. Thus one expects that entanglement disappears at large $t_{d}$. In this case we evaluate $E_c$ by measuring ($q_1,u_1$) at a time  delayed by  $t_d$ with respect ($q_2,u_2$) We plot in Fig.\ref{entanglement_versus_delay}a) the $E_c(t_d)$ as a function of $t_{d}$. At $t_{d}=0$ the entanglement coefficient is $0.85$ (i.e. the value plotted in Fig.\ref{fig:entanglement}a) at small $\Delta$) but it increases as a function of $t_{d}$. It becomes larger than 1 at about $4 \tau$, meaning that for $t_{d}>\tau$ Eq.\ref{explong} is not satisfied anymore and entanglement is lost. We see that the experimental and numerical values are in agreement with the theoretical prediction derived in ref.\cite{supmat}.
	
	Finally we checked the dependence of entanglement in non isothermal conditions with $T_1\ne T_2$. We changed $T_1$ in the range $[88K-300K]$ keeping $T_2=300K$. The measured entanglement coefficient is plotted as a function $T_1$ in Fig.\ref{entanglement_versus_delay}b) with the theoretical prediction. We see that in this case the maximum entanglement (minimum $Ec$) is reached in isothermal conditions and that the dependence on $T_1$ is very weak.  {In other words, in contrast to thermodynamics uncertainty relations \cite{Barato2015},  $E_c$ is not a  sensible measure of out of equilibrium}.  

 {As a conclusion	we've conducted experiments in classical statistics 
	that reveal a classical analog of entanglement} between two interacting electric circuits (similar to Brownian particles), akin to continuous variable quantum entanglement. {We have also studied how the results are affected by the finite sampling and coarsening time and we have established a maximum value of $\Delta$ above which the effect cannot be observed}. Our results suggest that the temperature of the thermal bath behaves similarly to Planck's constant in determining the uncertainty relations governing coordinate and coarse-grained velocity changes in the classical system. The Brownian
		entanglement is a consequence of a coarse-grained description and disappears for a finer resolution of the
		Brownian motion.  This experimental observation  align with the  theoretical predictions  regarding Brownian entanglement.  Remarkably, we observe that the mathematical equations describing the quantum and classical  scenarios share identical forms, highlighting a striking coincidence. Finally let us point out  another important and relevant difference between classical and quantum entanglement. Indeed the very definition of the coarse-grained velocities difference  Eq.\ref{eq:uj}, involves a global measure of the two subsystems as they depend both on $q_1$ and $ q_2$. Instead  in the quantum case the $p_n$ can be collected via local
			measurements on the corresponding quantum subensemble. 	 {It is worth to mention that  other position-momentum uncertainties in classical systems have been recently  derived  \cite{singh2025} and it would be interesting to study their consequences in the context of classical entanglement.} 
		\vskip 0.5cm

{\bf acknowledgments}
SC thanks G.Ciliberto,R.Livi,T.Roscilde and S. Ruffo for useful discussion. 		
		
		
\newpage
\bibliographystyle{unsrturl}
\bibliography{brownian}

\begin{thebibliography}{10}
\bibitem{Comment_1}
to simplify notation we do not explicitly write the $(\bold{X},t)$ dependence.
\end{thebibliography}
\vfill



\section*{Supplementary material}

# Supplementary material - Experimental realization of entanglement between two Brownian particles

Theerthagiri L[1,2], S Ciliberto[3*]

[1]*Physics Division, School of Science and Technology,*
*University of Camerino, I-62032 Camerino (MC), Italy*
[2]*Department of Physics, University of Naples "Federico II", I-80126 Napoli, Italy and*
[3]*Univ Lyon, ENS de Lyon, Univ Claude Bernard Lyon 1,*
*CNRS, Laboratoire de Physique, F-69342 Lyon, France*[*]
(Dated: August 9, 2024)

## I. THE PROBABILITY DISTRIBUTION OF $q_1$ AND $q_2$

The Fokker-Planck equation for the dynamics of the coupled Brownian particles Eq.1) of the main text can be solved directly and the outcome is known to be given by the following two-dimensional Gaussian distribution.

$$P(q_1, q_2, t) = \frac{1}{2\pi\sqrt{\mathcal{D}_C}} \exp\left\{\left[-\frac{1}{2}\sum_{i,j=1}^{2} \mathcal{C}_{ij} q_i q_j\right]\right\} \quad (1)$$

where

$$\mathcal{C}_{ij} = \frac{\sigma_{ij}}{\mathcal{D}_C} \quad (2)$$

with $\mathcal{D}_C = \sigma_{11}\sigma_{22} - \sigma_{12}^2$ and $\sigma_{ij} = <q_i q_j>$. In isothermal conditions, $T = T_1 = T_2$, $P(q_1, q_2)$ becomes the equilibrium distribution, with $\mathcal{C}_{ii} = A_i/(k_B T)$ and $\mathcal{C}_{ij} = -K/(k_B T)$.

## II. THE COARSE GRAINED VELOCITY

The conditional probability $P(\mathbf{y}, t + \Delta | \mathbf{X}, t)$ is known to satisfy the following Fokker-Planck equation [1]:

$$\partial_t P(\mathbf{y}, t+\Delta|\mathbf{X}, t) = -\sum_n \partial_{y_n} \left[\frac{f_n(\mathbf{y})}{R} P(\mathbf{y}, t+\Delta|\mathbf{X}, t)\right] \\ + \sum_n \frac{T_n}{R} \partial_{y_n y_n}^2 P(\mathbf{y}, t+\Delta|\mathbf{X}, t), \quad (3)$$

where $f_n(\mathbf{X}) = -\partial_{x_n} U(\mathbf{X})$, with $U = A_1 x_1^2/2 + A_2 x_2^2/2 - K x_1 x_2$ the potential energy of the system. The solution of Eq.3 can be written as:

$$P(\mathbf{y}, t+\Delta|\mathbf{X}, t) = \delta(\mathbf{X} - \mathbf{y})+ \\ + \frac{\Delta}{R} \sum_n [-f_n(\mathbf{X})\partial_{y_n} + T_n \partial_{y_n y_n}^2] \delta(\mathbf{X} - \mathbf{y}) + O(\Delta^2) \quad (4)$$

Inserting this solution in Eq.8) of the main text, the calculation of $V_{+,n}(\mathbf{X}, t)$ is straightforward upon using natural boundary conditions at infinity, i.e. $P(x_n, t) \to 0$ if $x_n \to \pm\infty$.

For $V_{-,n}(\mathbf{X}, t)$ we first rewrite Eq.8) as

$$V_{-,n}(\mathbf{X}, t, \Delta) = \int d\mathbf{y} \, \frac{(x_n - y_n)}{\Delta} \, P(\mathbf{y}, t-\Delta|\mathbf{X}, t) = \\ = \int d\mathbf{y} \, \frac{(x_n - y_n)}{\Delta} \, P(\mathbf{y}, t|\mathbf{X}, t-\Delta) \frac{P(\mathbf{y}, t-\Delta)}{P(\mathbf{X}, t)} \quad (5)$$

inserting Eq.4 in this equation and performing the limit $\Delta \to 0$ we get the result for $V_{-,n}(\mathbf{X}, t)$ in Eq.10 of the main text.

---

[*] sergio.ciliberto@ens-lyon.fr

### III. THE COARSE GRAINED VELOCITY DIFFERENCE

Eq. (1) and Eqs.9-11 of the main text produce for the coarse-grained velocity difference of the first and the second particle, respectively

$$u_1(\mathbf{X},t) = \frac{k_B T_1}{R} \frac{(\sigma_{22} q_1 - \sigma_{12} q_2)}{\mathcal{D}_C} \quad (6)$$

$$u_2(\mathbf{X},t) = \frac{k_B T_2}{R} \frac{(\sigma_{11} q_2 - \sigma_{12} q_1)}{\mathcal{D}_C} \quad (7)$$

The values of the covariances in the general case for the experimental set up are given in the next section. In the isothermal conditions $T = T_1 = T_2$ the coarse grained velocities simplify to:

$$u_1(\mathbf{X},t) = \frac{k_B T}{R} (A_1 \ q_1 - K \ q_2) \quad (8)$$

$$u_2(\mathbf{X},t) = \frac{k_B T}{R} (A_2 \ q_2 - K \ q_1) \quad (9)$$

### IV. DIRECT EVALUATION OF THE COARSE GRAINED VELOCITIES

The expression of $u_i$ in the previous section are estimated for $\Delta \to 0$. As explained in the main text we measured the dependence of $V_\pm$ as a function of $\Delta$. In order to perform such a measure we proceed in the following way. We find the maximum $q_{max}$ of $|q_1|$ and $|q_2|$ in all the data set. We define a vector $x_q$ of $n_q + 1$ components, such that $x_q(j) = -q_{max} + (j-1)dx_q$, with $j = 1,...,n_q+1$ and $dx_q = 2q_{max}/n_q$. Consider an $x_q(i)$ and find in the time series of $q_1$ all the times $\tilde{t}$ for which $x_q(i) - dx_q/2 < q_1(\tilde{t}) < x_q(i) + dx_q/2$. Then find in the time series $q_2(\tilde{t})$ all the times $\hat{t}$ for which $x_q(j) - dx_q/2 < q_2(\hat{t}) < x_q(j) + dx_q/2$. By definition the CGVD are given by :

$$u_n(xq(i), xq(j)) = < [-q_n(\hat{t}+\Delta) + 2 q_n(\hat{t}) - q_n(\hat{t}-\Delta)] >_{\hat{t}} /(2\Delta) \quad (10)$$

where $n = 1, 2$ and $< ... >_{\hat{t}}$ stands for the mean on all the times $\hat{t}$. The mean velocity is given by

$$\bar{V}_n(xq(i), xq(j)) = < [q_n(\hat{t}+\Delta) - q_n(\hat{t}-\Delta)] >_{\hat{t}} /(2\Delta) \quad (11)$$

### V. THE COVARIANCES FOR THE EXPERIMENT

For the experiment of Fig.1 and Eq.1 of the main text one obtains for the variances of $q_i$ (see Ref.[2]) :

$$\sigma_{11} = < q_1^2 > = k_B T_1 (C_1 + C) + \dot{Q} \ R \ D \quad (12)$$

$$\sigma_{22} = < q_2^2 > = k_B T_2 (C_2 + C) - \dot{Q} \ R \ D \quad (13)$$

with the heat flux $\dot{Q}$ between the two heat baths given by

$$\dot{Q} = \frac{k_B (T_2 - T_1) C^2}{D \ Y} \quad \text{with} \quad Y = R(C_2 + C_1 + 2C) \quad (14)$$

. The covariance is :

$$\sigma_{12} = < q_1 q_2) > = \frac{k_B C \ R}{Y}[T_1(C_1 + C) + T_2(C_2 + C)] \quad (15)$$

From the values of $\sigma_{ij}$ one gets:

$$\mathcal{D}_C = k_B^2 D T_1 T_2 + \dot{Q}^2 D^3 R^2 / C^2 \quad (16)$$

Using Eqs.12-16 in Eqs.6,7, we obtain for the coarse grained velocity covariances

$$< u_1^2 > = (k_B T_1 / R)^2 \sigma_{22} / \mathcal{D}_C \quad (17)$$

$$< u_2^2 > = (k_B T_2 / R)^2 \sigma_{11} / \mathcal{D}_C \quad (18)$$

$$< u_1 u_2 > = -(k_B / R)^2 T_1 T_2 \sigma_{12} / \mathcal{D}_C . \quad (19)$$

## VI. THE ENTANGLEMENT COEFFICIENT

The entanglement coefficient $E_c$ is the ratio between the left and the right hand sides of Eq.16 of the main text. In isothermal condition $T = T_1 = T_2$ and when the measurements of $q_1$ and $q_2$ are delayed by a time $t_d$, then it takes the form :

$$E_c = \frac{1}{4k_B T}[<q_1^2> A_1 + <q_2^2> A_2 + $$
$$R^2 \frac{<u_1^2> A_2 + <u_2^2> A_1}{A_1 A_2} + $$
$$- 2 <q_1(t_d) q_2(0)> \sqrt{A_1 A_2}$$
$$+ 2 \frac{<u_1(t_d) u_2(0)> R^2}{\sqrt{A_1 A_2}}] \qquad (20)$$

where we used the values $\epsilon = -1$ and $\zeta = 1$ which maximize entanglement. Inserting in Eq.20 the expression of correlation functions computed in sec.VII one gets :

$$E_c(t_d) = 1 + \frac{C^2}{2D} - \frac{C \exp(-a|t_d|)}{2\sqrt{D+C^2}}\left[\left(2 + \frac{C^2}{D}\right)\cosh(\Psi|t_d|) + \frac{a\ C^2}{\Psi\ D}\sinh(\Psi|t_d|)\right] \qquad (21)$$

with $a = (C_1 + C_2 + 2C)/(2DR)$ and $\Psi = \sqrt{a^2 - 1/(R^2 D)}$.
This theoretical result is compared with the experimental one in Fig.5 of the main text.
At $t_d = 0$ it is :

$$E_c = 1 + \frac{C^2}{2D} - \frac{C}{2\sqrt{D+C^2}}\left(2 + \frac{C^2}{D}\right) \qquad (22)$$

For the experimental parameter $E_c = 0.855$ which is the black horizontal dashed line in Fig.4 of the main text.

## VII. CORRELATION FUNCTIONS

$$\langle q_1(t_d) q_2(0)\rangle = k_B T\ C\ \exp(-a|t_d|)\left[\frac{a}{\Psi}\sinh(\Psi|t_d|) + + \cosh(\Psi|t_d|)\right] \qquad (23)$$

$$\langle q_1(t_d) q_1(0)\rangle = k_B T(C_1 + C)\exp(-a|t_d|)\left[\cosh(\Psi|t_d|) + \frac{(C_1+C)^2 + C^2 - D}{2(C_1+C)\ R\ D\ \Psi}\sinh(\Psi|t_d|)\right] \qquad (24)$$

$$\langle q_2(t_d) q_2(0)\rangle = k_B T(C_2 + C)\exp(-a|t_d|)\left[\cosh(\Psi|t_d|) + \frac{(C_2+C)^2 + C^2 - D}{2(C_2+C)\ R\ D\ \Psi}\sinh(\Psi|t_d|)\right] \qquad (25)$$

$$\langle u_1(t_d) u_2(0)\rangle = \frac{(X+2C^2)}{R^2 X^2} <q_1(t_d)q_2(0)> - \frac{C(C_1+C)}{R^2 D^2} <q_2(t_d)q_2(0)> - \frac{C(C_2+C)}{R^2 D^2} <q_1(t_d)q_1(0)> \qquad (26)$$

with

15

$$a = \frac{(C_1 + C_2 + 2C)}{2DR} \quad \text{and} \quad \Psi = \sqrt{a^2 - \frac{1}{R^2 D}} \qquad (27)$$

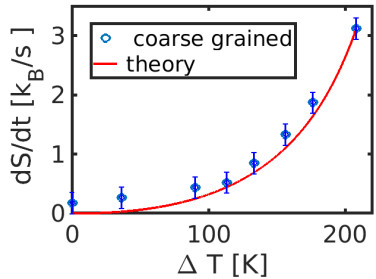

FIG. 1. Entropy production versus different temperature. The experimental points have been obtained from the measured coarse grained velocities in left hand side of eq.31. The red line has been computed inserting in the definition of $\dot Q$, Eq.14, the experimental values of the parameters.

## VIII. ENTROPY PRODUCTION AND COARSE GRAINED VELOCITY

The mean coarse grained velocity is defined in the main text as

$$\bar V_n(\mathbf{X}) = \frac{V_{-,n}(\mathbf{X}) + V_{+,n}(\mathbf{X}))}{2} \tag{28}$$

which using Eqs.9),10 and 11) of the main text can be rewritten as :

$$\bar V_n(\mathbf{X}) = \frac{f_n(\mathbf{X})}{R} - \frac{k_B T_n}{R}\partial_{x_n}\ln P(\mathbf{X}) \tag{29}$$

$$= \frac{1}{R}(-A_n q_n + K q_{n'} - u_n R) \quad \text{with} \quad n' \neq n, \tag{30}$$

Using uncertainty relations $\langle u_n q_{n'}\rangle = \delta_{n,n'} k_B T_n/R$, the covariances displayed in Sec.V, Eqs.6,7 and the fact that

$$-A_n + \langle u_n^2\rangle R^2/T_n = A_{n'} - \langle u_{n'}^2\rangle R^2/T_{n'},$$

after some algebra we get the entropy production rate :

$$\dot Q\left(\frac{1}{T_1} - \frac{1}{T_2}\right) = R\left(\frac{\bar V_1(\mathbf{X})^2}{T_1} + \frac{\bar V_2(\mathbf{X})^2}{T_2}\right) \tag{31}$$

where $\dot Q$ is the heat flux between the two bath defined in Eq.14 [2]. In Fig.1 we plot the experimental and theoretical values of the entropy production rate as a function of $\Delta T = T_2 - T_1$. The experiment agrees wthin error bars with the theoretical prediction which is obtained using $\dot Q$ computed directly from Eq.14. The experimental points are obtained using in the right hand side of Eq.31 the measured values of $\bar V_n$. Thus Eq.31 strengths the physical meaning of the coarse grained velocities fluctuations.

---


\end{document}